# Two-Photon Emission from Semiconductors

**Alex Hayat, Pavel Ginzburg and Meir Orenstein**

Department of Electrical Engineering, Technion, Haifa 32000, Israel

We report the first experimental observations of two-photon emission from semiconductors, to the best of our knowledge, and develop a corresponding theory for the room-temperature process. Spontaneous two-photon emission is demonstrated in optically-pumped bulk GaAs and in electrically-driven GaInP/AlGaInP quantum wells. Singly-stimulated two-photon emission measurements demonstrate the theoretically predicted two-photon optical gain in semiconductors - a necessary ingredient for any realizations of future two-photon semiconductor lasers. Photon-coincidence experiment validates the simultaneity of the electrically-driven GaInP/AlGaInP two-photon emission, limited only by detector's temporal resolution.

Two-photon emission (TPE) is a process in which electron transition between quantum levels occurs via simultaneous emission of two photons. This phenomenon is important for astrophysics and atomic physics [1, 2], while semiconductor TPE was recently proposed as a compact source of entangled photons, essential for practical quantum information processing [3-5], three orders of magnitude more efficient[6] than the existing down-conversion schemes. Two-photon absorption in semiconductors has been substantially investigated[7-11]; however spontaneous semiconductor TPE has been neither observed nor fully analyzed theoretically so far.

Two-photon transition is basically much weaker than the related first-order process. Therefore observations of multi-photon spontaneous decays have so far been restricted to a few atomic transition cases, where the lowest-order transition is forbidden by selection rules[16] or suppressed by a cavity[17], while their two-photon spectrum is continuous and centered at about half the one-photon transition frequency [1,18]. Semiconductors can be injected with very high charge carrier densities, making even the weak second-order spontaneous processes measurable, and their TPE spectrum is expected to be determined by the photonic state density as well as by the carrier energy distribution.

Here we report the first experimental observation of spontaneous TPE from room-temperature semiconductors, excited both optically in bulk GaAs and electrically in GaInP/AlGaInP quantum wells (QWs) grown on GaAs substrate. We develop a theoretical model for spontaneous TPE from semiconductors validating the measured results. In additional experiments, the optically pumped bulk GaAs was singly-stimulated to emit specific photon pairs by launching photons at a prescribed wavelength – yielding the co-emission of complementary wavelength photons with the suppression of the rest of the spectrum. Similar results are shown for GaInP/AlGaInP QWs, while photon coincidence experiment verifies the simultaneity of TPE.

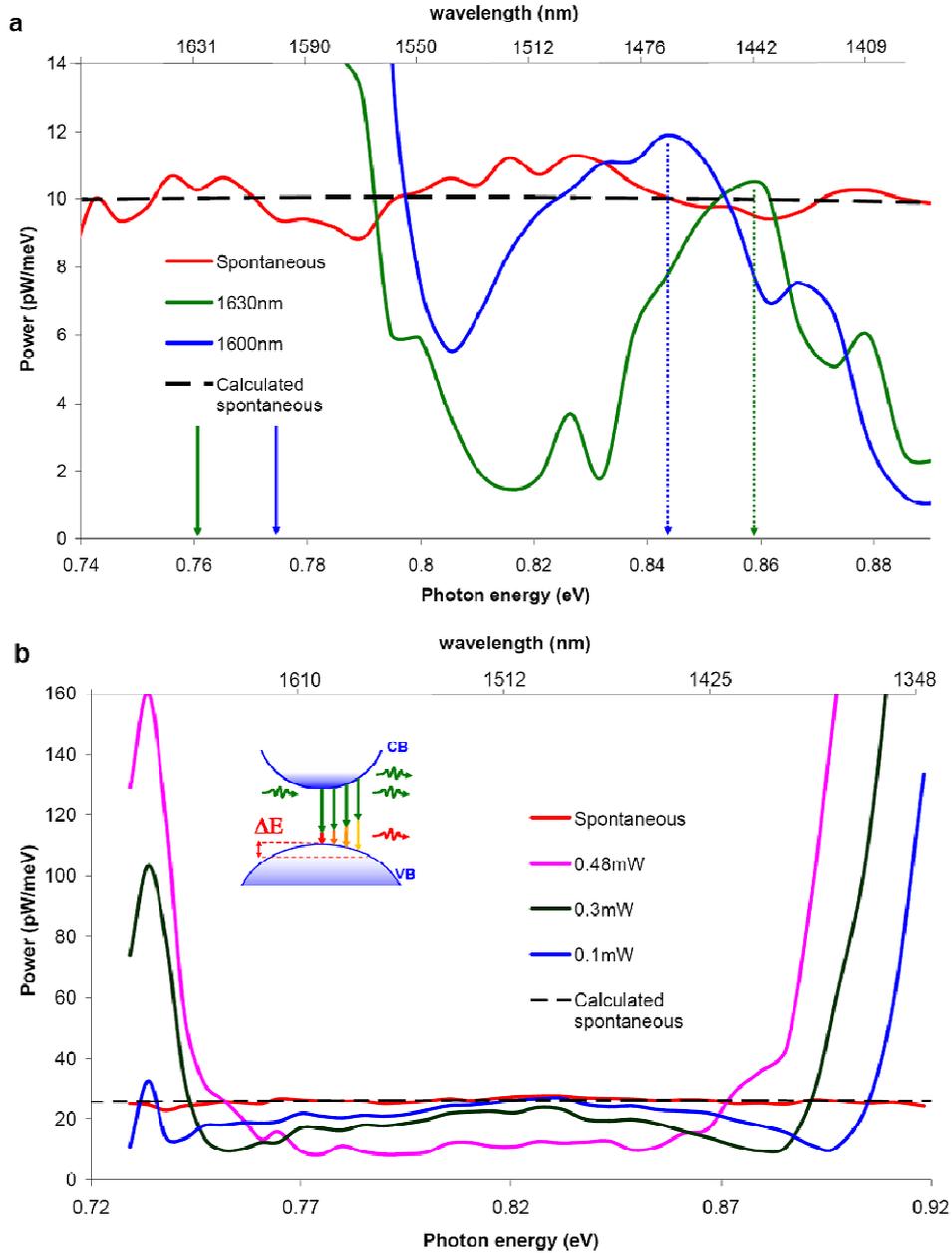

**Fig. 1**. Bulk GaAs TPE measurements and calculations with optical pumping by 514nm Ar laser: **(a)** 100mW pump spontaneous and singly-stimulated TPE at different wavelengths with 0.2mW stimulation power. The solid arrows indicate the stimulating photon energies and the dashed arrows indicate the theoretical stimulated photon central energies. **(b)** 180mW pump spontaneous and singly-stimulated TPE with various stimulation powers at 1310nm. The inset is a schematic description of singly-stimulated TPE energy broadening in bulk semiconductor.

In our first experiment, we optically pumped GaAs by a 100mW continuous wave (CW) 514nm argon laser ($\hbar\omega_{pump} \approx 1.73 E_{gap}$), focused with ~30μm spot size (after filtering out spontaneous infrared emission) onto a ~200μm thick GaAs sample (much thicker than the laser penetration depth), inducing ~$1.2 \cdot 10^{18}$cm$^{-3}$ local carrier concentration, while the pump-induced sample heating is estimated to be ~330K. The pump laser was chopped at 236Hz, and the collected emission from the sample in a transmission configuration was detected by a New-Focus infrared (IR) femtowatt photo-receiver via a lock-in amplifier. The spectrum was obtained by an Acton-Research monochromator using a 1600nm-blazed grating and ~10nm spectral resolution. The measured spectrum for the TPE was very wide as expected (Fig. 1-a) complying with the theory described below, while the overall collected optical power was ~3nW.

In order to validate unambiguously that the observed luminescence is indeed a TPE, and to dismiss any possibility of inhomogeneously broadened emission from mid-gap levels, we measured singly-stimulated TPE by launching a specific wavelength into the optically pumped GaAs. As a result of this excitation, a complementary TPE wavelength peak appeared. In all the stimulated TPE experiments only the carrier density was modulated for lock-in detection, whereas the stimulating lasers were not modulated, while the observed signals were proportional to the pump power. Stimulation with 0.2mW 1630nm (0.761eV) laser resulted in a peak at 1451nm (0.854eV), while a 1600nm (0.775eV) laser yielded a peak at 1476nm (0.84eV) (Fig.1-a). In both cases the photon energies are complementary about the center of the spontaneous emission at 0.81eV within a ~3meV error, as expected due to energy conservation. Changing the stimulation wavelength to 1570nm (0.79eV) caused the stimulating and the complementary peaks to merge approaching the degenerate case. The peaks appeared to merge for any stimulation energy closer than 30meV to the spontaneous TPE center. Furthermore, during this excitation, the rest of the spontaneous emission spectrum was suppressed (Fig. 1), clearly demonstrating that the wideband emission is homogeneously broadened. In TPE, the wideband homogeneous broadening stems from the very short lifetime of the intermediate virtual state, in contrast to an inhomogeneously broadened emission from any possible deep levels within the band-gap[19].

In a third experiment, we have performed singly-stimulated TPE measurements in order to study the dependence of the complementary wavelength peak on the stimulating power. To increase the dynamic range of the measurement the pump power was increased to 180mW resulting in ~2·$10^{18}$ $cm^{-3}$ carrier concentration and a 0.84eV TPE center. Stimulation with 1310nm (0.946eV) laser yielded a complementary peak at 1691nm (0.733eV), while the observed power-dependence is shown to be linear (Fig1 b), as expected from the theory, where the complementary peak width ΔE is related to the carrier energy distribution (Fig1-b inset).

Hence, due to the observations described above, the measured IR luminescence is validated as TPE from bulk GaAs. These stimulated TPE experiments also demonstrate two-photon optical gain in semiconductors, which was proposed theoretically in a degenerate case for two-photon semiconductor amplifiers[12, 14, 15] and two-photon semiconductor lasers[13]. Two-photon gain and lasing were observed previously only in discrete-level atomic systems[20, 21] exhibiting interesting nonlinear behavior. Usually two-photon lasers require seed optical power to achieve threshold. Given the nonlinearity of two-photon gain, despite the fact that one-photon oscillator strength is much larger, at high seed intensities two-photon gain can become the dominant process. Moreover, semiconductor laser complex cavities can be designed for the two-photon wavelengths only, which do not support any one-photon wavelength modes, considering the material dispersion or specially designed reflectors. Employing this effect in semiconductor materials may result in novel integrated-photonics sources allowing ultra-short pulse generation[15], and nonlinear devices for optical switching and digital logic[22].

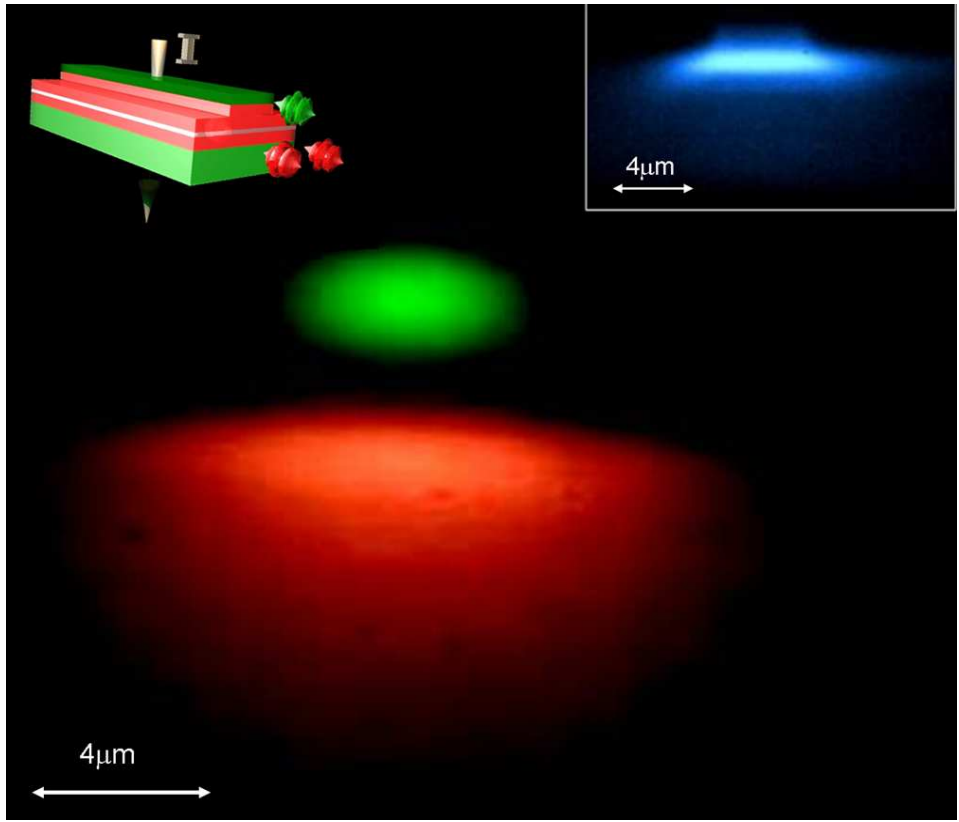

**Fig. 2.** Experimental false color IR emission imaging of the facet of the GaInP/AlGaInP QWs based waveguide. The upper lobe (green) is the ~1.4eV one-photon emission from the GaAs cap layer and the lower lobe (red) is the wide-band TPE from GaInP/AlGaInP QWs at ~200mA injection current. The inset is a snapshot of the visible emission from the same structure at ~10mA injection current.

In a fourth experiment, we made a further step towards the device implementation of the effect. In this electrically driven TPE experiment we employed semiconductor samples consisting of 4 periods of compressively strained 50Å $Ga_{0.45}In_{0.55}P$ QWs separated by 55Å $(Al_{0.5}Ga_{0.5})_{0.51}In_{0.49}P$ barriers, ~1.1μm AlGaInP cladding and a heavily doped 300nm GaAs cap layer. The lateral light confinement was achieved by a 4μm-wide ridge waveguide, realized by etching techniques, in order to enhance the emitted photons collection efficiency. The structure was electrically pumped below the lasing threshold, which was raised to above 200mA by applying an anti-reflection coating on the facets, in order to enable higher carrier population inversion, while the samples were kept at 300K by closed-loop thermoelectric cooling. Spatial distribution of the wide-band IR TPE intensity was centered near the QWs, whereas the one-photon ~1.4eV narrow-band emission was located in the GaAs cap layer region (Fig. 2). Carrier population was modulated by the injection current at 316Hz for the lock-in detection through a 500nm-blazed grating monochromator. The measured spectrum for the TPE was very wide as expected and centered at 0.98eV in good agreement with the calculations (Fig. 3-a) using theoretical model described below. The overall TPE measured optical power was around 30nW, which is 5 orders of magnitude weaker than the measured one-photon emission (Fig. 3-a inset), corresponding to TPE rates much higher than TPA at a single-photon level due to a continuum of photonic vacuum states in TPE.

The fifth experiment was aimed at observing singly-stimulated TPE in electrically-pumped GaInP/AlGaInP QWs. The stimulating laser was facet-coupled into the waveguide with about 5μm$^2$ mode area by a polarization maintaining lensed fiber (Fig. 3-b inset). The measurements were taken using a 500nm-blazed grating 5nm-resolution monochromator. Stimulation with 1500nm (0.826eV) laser resulted in a complementary peak at 1100nm (1.127eV), and 1450nm (0.855eV) stimulation yielded a 1117nm (1.11eV) peak (Fig. 3-b). Both cases are complementary around the center of TPE at 0.98eV according to the theory within 4meV error. The width of the complementary peak is close to the one-photon emission spectral width, as expected (Fig 3-a inset), however the effect of stimulation is much smaller in QWs than in bulk GaAs due to the very small overlap of the optical mode with the QWs (<10$^{-2}$).

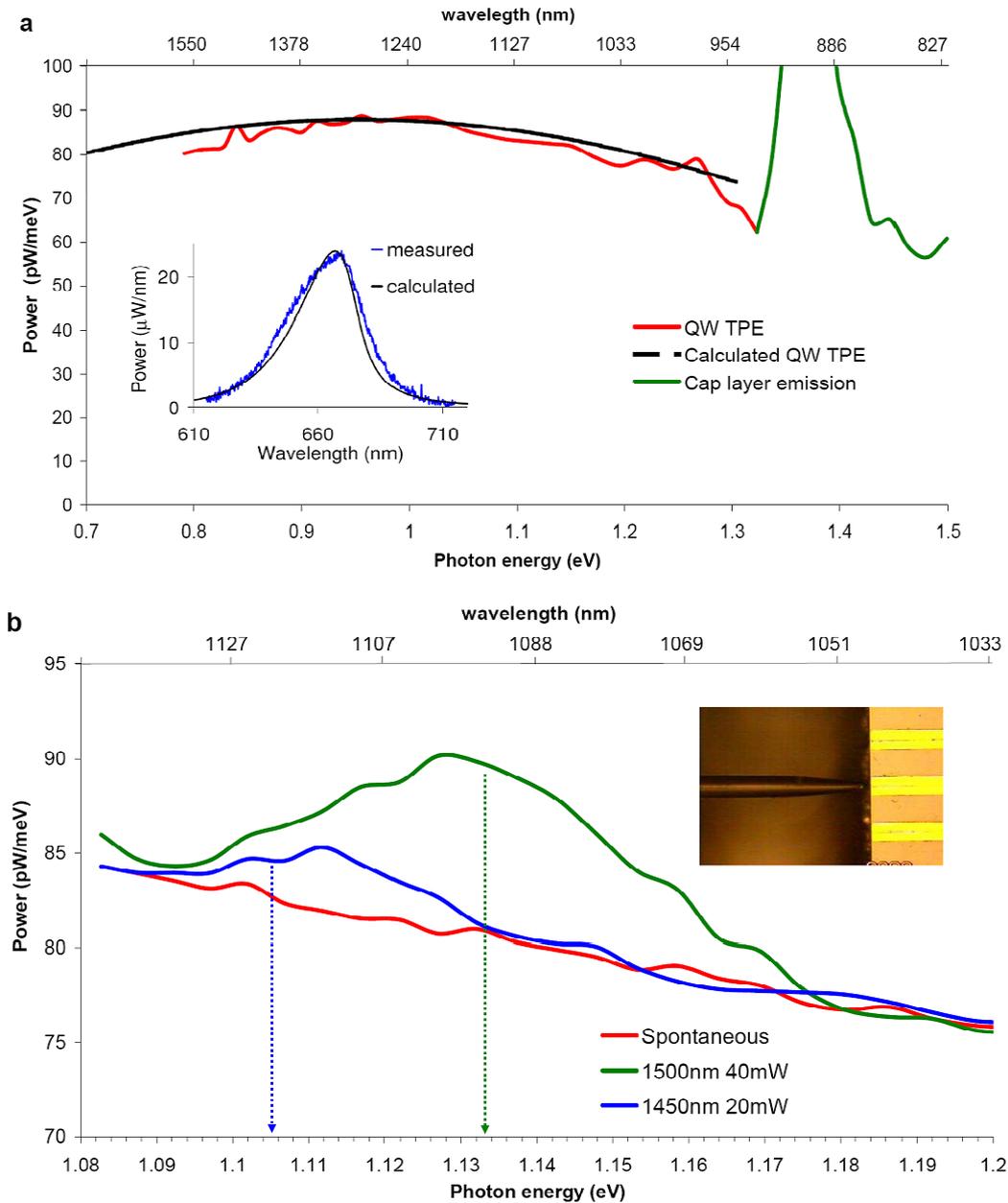

**Fig. 3**. Measured and calculated IR emission spectrum from GaInP/AlGaInP QWs at ~200mA injection current. **(a)** Spontaneous TPE. The narrow-band ~1.4eV peak (green) is the one-photon emission from the GaAs cap layer and the wide-band TPE (red) is from the GaInP/AlGaInP QWs. The inset is the measured and calculated one-photon emission spectrum from GaInP/AlGaInP QWs at ~200mA injection current. **(b)** Singly-stimulated TPE measurement from GaInP/AlGaInP QWs with different stimulation wavelengths and powers. The dashed arrows indicate the theoretical stimulated photon central energies.

In developing semiconductor TPE theory, a fundamental difference between two-photon processes in semiconductors and in discrete-level atomic systems that must be considered is the solid-state delocalized electron wavefunctions with definite crystal momentum in contrast to the localized atomic states. The rate calculations for TPE require summing over a complete set of the system states[1]. Continuum of delocalized electron states in every band of the semiconductor introduces a vast number of close intermediate states, making semiconductor two-photon process more efficient. Furthermore, using intra-band transitions as part of the two-photon second-order calculations results in a k-dependence of the matrix element unlike for the inter-band transitions for parabolic bands. Spontaneous TPE spectrum from a semiconductor can be calculated similar to atomic scenarios[23] by a second order time-dependent perturbation term using the $H = -\frac{e}{m_0}\hat{\vec{A}} \cdot \hat{\vec{p}}$ electron-radiation interaction Hamiltonian in the dipole approximation, where $\hat{\vec{p}}$ is the momentum operator, $e$ is the electron charge, $m_0$ – free space electron mass and $\hat{\vec{A}}$ is the vector potential operator for a lossless dielectric in Coulomb gauge given by:

$$\vec{A}(r,t) = \sum_{\vec{q}} \sqrt{\frac{\hbar}{2 n_q^2 \varepsilon_0 \omega_q V_c}} \hat{\varepsilon}_q \left( \hat{a}_q^+ e^{i\omega_q t - i\vec{q}\cdot\vec{r}} + \hat{a}_q e^{-i\omega_q t + i\vec{q}\cdot\vec{r}} \right) \quad (1)$$

with $\hat{\varepsilon}_q$ – photon polarization, $\varepsilon_0$ - vacuum permittivity, $\vec{q}$ – photon wavevector, $\omega_q$ – photon radial frequency, $V_c$ – field quantization volume, $\hat{a}_q$ and $\hat{a}_q^+$ are the field annihilation and creation operators and $n_q$ - material refractive index at $\omega_q$. Using the photon state density per unit frequency of $\rho_p(\omega_i) = \dfrac{V_c \omega_i^2 n_i^3 d\Omega_i}{(2\pi)^3 c^3}$ for each of the two emitted photons into a solid angle $d\Omega_i$ with $c$ - vacuum light speed, TPE rate in bulk semiconductor and in QWs is calculated per unit frequency:

$$\left.\frac{W}{\partial \omega_1}\right|_{Bulk} = \int dk \, |\vec{k}|^4 V_{Bulk} M^2(\omega_1, E_{21})$$

$$\left.\frac{W}{\partial \omega_1}\right|_{QW} = \int dk \, |\vec{k}|^3 \pi |\langle \phi_c | \phi_v \rangle|^2 S_{QW} M^2(\omega_1, E_{21})$$

(2)

with the dimensionless matrix element $M$, calculated considering the initial and final states as intermediate states, similar to semiconductor TPA calculations[9]:

$$M^2(\omega_1, E_{21}) = \left(\frac{e}{m_0}\right)^4 \frac{p_{cv}^2 n_1 n_2}{2\pi^5 \varepsilon_0^2 c^6} \left(\frac{m_0}{m_r}\right)^2 \left|\frac{1}{i\omega_1 + \Gamma} + \frac{1}{i(E_{21}/\hbar - \omega_1) + \Gamma}\right|^2 \times$$
$$\times f_2(|\vec{k}|)(1 - f_1(|\vec{k}|)) \omega_1 (E_{21}/\hbar - \omega_1)$$

(3)

where $m_r$ is the reduced electron mass, $p_{cv}$ is the bulk material momentum matrix element[8], $E_{21}$ - electron-hole energy separation, $\vec{k}$ - corresponding electron crystal momentum, $V_{Bulk}$ - bulk semiconductor sample volume, $S_{QW}$ - QW sample area, $f_1$ and $f_2$ - conduction band (CB) and valence band (VB) Fermi-Dirac population functions, $\phi_c$ and $\phi_v$ - CB and VB QW ground state envelope functions, and $\Gamma$ is a phenomenologically included transition dephasing [24, 25].

TPA calculations performed without this dephasing term were confirmed by experiments[7,8]. Nevertheless, all non-degenerate TPA calculations[10, 11] were done far from the extreme case of $\hbar\omega_1 \to 0$ and $\hbar\omega_2 \to E_{21}$, where one-photon absorption is dominant. In spontaneous TPE calculations, however, a complete continuum of frequencies must be considered, and in the absence of the dephasing term $\Gamma$, the matrix element divergence for $\omega_1 \to 0$ or $(E_{21}/\hbar - \omega_1) \to 0$ (Eq. 3) causes an infrared catastrophe similar to those occurring in zero-frequency nonlinear optics calculations[26].

TPE spectra from bulk GaAs at 330K for the discussed injection levels were calculated according to the described model, using the CB and VB as the intermediate states as well, while neglecting other intermediate bands due to their energy remoteness[8,27] using the Joyce-Dixon approximation[28] and including bandgap shrinkage effects[29] (Fig 1), where the transition dephasing $\Gamma$ is determined mainly by the carrier decoherence time - less than 100fs for such injection levels. GaInP/AlGaInP QWs TPE spectrum was also calculated for the given current density (Fig 3-a). According our model, the ratio of spontaneous TPE versus one-photon emission is in general dependent on the injection levels because the TPE matrix element is k-dependent while one-photon emission is not (for parabolic bands). However, for the specific injection levels considered in GaInP/AlGaInP QWs, TPE is 5 orders of magnitude weaker than one-photon emission. This ratio is in good agreement with the experiments presented.

In the sixth and last experiment, the photon emission simultaneity was demonstrated by probing the coincidences in TPE from GaInP/AlGaInP QWs. In semiconductors, TPE coincidence measurements are more involved than in typical atomic systems due to the relatively low photon energies. For GaInP/AlGaInP QW TPE, at least one InGaAs-based photon counter (PC) must be used. Unlike Si PCs, InGaAs-based PCs must be operated in gated mode due to high dark-count rates, and they suffer from significant afterpulsing[30] (~50μs trapped-carrier lifetime at ~200K). In our GaInP/AlGaInP QW TPE coincidence experiment, one Si PC (Perkin Elmer) and one

InGaAs PC (Princeton Lightwave) cooled to ~200K were used. The sample was driven by 10ns current pulses at 100KHz repetition rate, which were synchronized with the InGaAs PC gating. Pulse magnitude was set to ~1000 PC counts/sec. For this pulsed operation, delays between PC's outputs different from an integer number of periods (10μs) resulted in vanishing correlations. Zero relative delay yielded near 10% coincidences due to the optical components losses. Negative integer-period number relative delay (Si output before InGaAs) resulted in about 0.5% coincidences, close to the calculated accidental background level (Fig 4). Positive integer-period number relative delay (Si output after InGaAs) resulted in coincidences rates decaying nearly exponentially with ~50μs time constant due to the InGaAs PC afterpulsing.

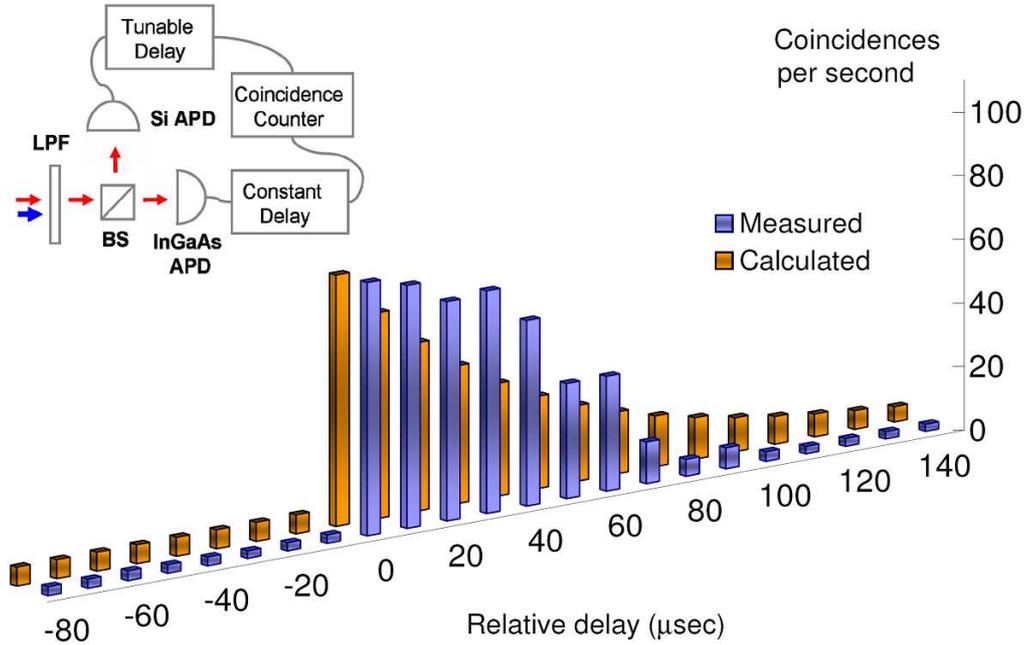

**Fig. 4.** Calculated and measured photon coincidences in electrically pumped GaInP/AlGaInP QWs TPE vs relative delay between detectors. The inset is a schematic setup description.

In conclusion, we have demonstrated experimentally and analyzed theoretically TPE from semiconductors, which has not been studied before. Spontaneous and singly-stimulated TPE was demonstrated in optically-pumped GaAs and in current-driven GaInP/AlGaInP QWs. Our theoretical calculations validate the experimental observations. The overall TPE is relatively strong – around 30nW in GaInP/AlGaInP QWs. Such high TPE intensities may help exploiting this fundamental physical phenomenon for practical uses in both science and engineering, while the demonstrated photon coincidences are crucial for quantum applications such as photon entanglement.